\documentstyle[preprint,tighten,aps,epsfig]{revtex}
\begin{document}
\draft
\preprint{
\begin{tabular}{r}
DFTT 6/96
\\
SISSA Ref.28/96/EP
\\
hep-ph/9602383
\\
\end{tabular}
}
\title{Possible tests of
neutrino maximal mixing
and comments on matter effects}
\author{S.M. Bilenky}
\address{Joint Institute for Nuclear Research,
Dubna, Russia,
and
\\
SISSA-ISAS, Trieste, Italy.}
\author{C. Giunti}
\address{INFN,
Sezione di Torino and Dipartimento di Fisica Teorica,
Universit\`a di Torino,
\\
Via P. Giuria 1, 10125 Torino, Italy.}
\author{C. W. Kim}
\address{Department of Physics and Astronomy,
The Johns Hopkins University,
\\
Baltimore, Maryland 21218, USA.}
\date{\today}
\maketitle
\begin{abstract}
We show in a simple and general way
that matter effects
do not contribute
to the averaged value of the
probabilities of transition
of solar $\nu_e$'s into other states
in the case of maximal mixing of
any number of massive neutrinos.
We also show
that future solar neutrino experiments
(Super-Kamiokande and SNO)
will allow to test
the model with maximal mixing of three
massive neutrinos
in a way that does not depend
on the initial solar neutrino flux.
\end{abstract}

\pacs{}

\narrowtext

The hypothesis of maximal neutrino mixing
was proposed
many years ago
\cite{Gribov69,Nussinov76,Wolfenstein78}.
Recently
the interest in this hypothesis
is increased
in connection with
the atmospheric neutrino anomaly
\cite{KAM94,IMB,SOUDAN},
whose explanation requires
neutrino oscillations
with large mixing.
There are some theoretical arguments
in favor of maximal mixing
based on the see saw mechanism
\cite{Smirnov93,CWKim93}
and on the hypothesis of permutation symmetry
between generations at the unification scale
\cite{HS94}. 
In Refs.\cite{HPS95,GKK95}
the data of the experiments on the detection of
solar and atmospheric neutrinos
were analyzed
under the assumption
of maximal mixing of three
neutrino fields
with masses
$ m_1 \ll m_2 \ll m_3 $.
It was shown that the
atmospheric neutrino data are well described
in this model with
$ \Delta m^2_{31} \simeq 10^{-2} \, \mbox{eV}^2 $.
The data of all four solar neutrino experiments
\cite{HOMESTAKE,KAMIOKANDE,GALLEX,SAGE}
cannot be described in this model
with an acceptable value of $\chi^2$.
If only the data of
the Kamiokande, GALLEX and SAGE experiments
\cite{KAMIOKANDE,GALLEX,SAGE}
are taken into account,
it is possible to obtain a good fit
in the case of maximal mixing
with
$ \Delta m^2_{21} \lesssim 10^{-12} \, \mbox{eV}^2 $.

In the recent paper
\cite{HPS96}
it was shown that in the case of maximal mixing
matter effects do not modify the 
vacuum oscillation probability
of solar $\nu_e$'s to survive.
This was shown with a numerical
solution of the evolution equation of neutrinos in matter
for the case of two and three generations.
In the first part of this note
we will show in a simple and general
way that
the averaged transition probabilities
of solar $\nu_e$'s into any
final state
do not depend on the presence of matter
in the case of the maximal mixing.

Let us consider first the
very well known case
of vacuum oscillations
(see, for example, Ref.\cite{Bilenky}).
The averaged probability of the transition of solar
$\nu_e$'s into any state
$\nu_{\alpha}$ in the general case of $N$
massive neutrinos is given by
\begin{equation}
\left\langle P_{\nu_e\to\nu_\alpha} \right\rangle
=
\sum_{j=1}^{N}
\left| U_{{\alpha}j} \right|^2
\,
\left| U_{ej} \right|^2
\;,
\label{51}
\end{equation}
where
$U_{{\alpha}j}$
are the elements of the
mixing matrix.
In Eq.(\ref{51}) we have assumed that,
due to averaging
over the neutrino spectrum
and over the regions where neutrinos are
produced and detected,
all the
interference terms in the transition probabilities
disappear. 
This takes place if all the neutrino mass-squared differences
satisfy the following inequalities
\begin{equation}
\left| \Delta m^2_{jk} \right|
\gg
10^{-10} \, \mbox{eV}^2
\qquad \qquad
(j \not= k)
\;,
\label{101}
\end{equation}
where
$ \Delta m^2_{jk} \equiv m_j^2 - m_k^2 $.
In the case of maximal mixing
\cite{Gribov69,Nussinov76,Wolfenstein78}
\begin{equation}
\left| U_{{\alpha}j} \right|^2
=
{ 1 \over N }
\label{52}
\end{equation}
and we obtain the well known result
\cite{Pontecorvo71}
\begin{equation}
\left\langle P_{\nu_e\to\nu_\alpha} \right\rangle
=
{ 1 \over N }
\;.
\label{53}
\end{equation}
Let us notice that this is the minimum value for
the averaged probability of solar
$\nu_e$'s to survive.
Using this expression (in the case $N=2$)
many years ago B. Pontecorvo
\cite{Pontecorvo67}
for the first time noticed that
the detected flux of solar $\nu_e$'s can be as small as
half of the expected flux.

Let us now consider solar $\nu_e$'s
that are produced in
the core of the sun
with energy $E$.
As a result of neutrino propagation
in the interior of the sun
the neutrino state on the surface of the sun
has the following form
\begin{equation}
\left| \nu \right\rangle_{\odot}
=
\sum_{\alpha}
A^{e}_{\alpha}(E)
\left| \nu_{\alpha} \right\rangle
\;,
\label{11}
\end{equation}
where
$ \left| \nu_{\alpha} \right\rangle $
is any neutrino state
(active or sterile).
The amplitudes
$ A^{e}_{\alpha}(E) $
are determined by
the evolution equation of neutrinos
in matter.
The values of these amplitudes are not
important here.
We will use only the normalization condition
\begin{equation}
\sum_{\alpha}
\left| A^{e}_{\alpha}(E) \right|^2
=
1
\;.
\label{12}
\end{equation}
Let us stress that
the amplitudes
$ A^{e}_{\alpha} $
can depend on the neutrino energy $E$.

The neutrino state on the earth
is given by
\begin{equation}
\left| \nu \right\rangle_{\oplus}
=
\sum_{\beta,j,\alpha}
A^{e}_{\beta}
U_{{\beta}j}^{*}
\mbox{e}^{-i E_j T}
U_{{\alpha}j}
\left| \nu_{\alpha} \right\rangle
\;,
\label{111}
\end{equation}
where $ T \simeq R $,
$R$ being the distance between the
surface of the sun and the earth.

The probability of transitions of
solar $\nu_e$'s into any state
$\nu_{\alpha}$
is given by
\begin{equation}
P_{\nu_e\to\nu_\alpha}
=
\sum_{\beta,\rho,j,k}
A^{e}_{\beta}
\,
{A^{e}_{\rho}}^{*}
\,
{U_{{\beta}j}}^{*}
\,
U_{{\alpha}j}
\,
U_{{\rho}k}
\,
U_{{\alpha}k}^{*}
\,
\exp
\left(
- i
\,
{\displaystyle
\Delta m^2_{jk}
\over\displaystyle
2 E
}
\,
R
\right)
\;.
\label{13}
\end{equation}
If we assume that the squared mass differences
satisfy Eq.(\ref{101}),
all the interference terms disappear
in the expression for
the measurable
averaged probability,
which is given by
\begin{equation}
\left\langle P_{\nu_e\to\nu_\alpha} \right\rangle
=
\sum_{j=1}^{N}
\left| U_{{\alpha}j} \right|^2
\left|
\sum_{\beta}
A^{e}_{\beta}
\,
U_{{\beta}j}^{*}
\right|^2
\;.
\label{14}
\end{equation}
In the case of maximal mixing
$ \left| U_{{\alpha}j} \right|^2 = 1/N $
and from Eq.(\ref{14}),
using the unitarity of the mixing matrix
and the normalization condition (\ref{12}),
we obtain
\begin{equation}
\left\langle P_{\nu_e\to\nu_\alpha} \right\rangle
=
{ 1 \over N }
\sum_{j=1}^{N}
\left|
\sum_{\beta}
A^{e}_{\beta}
\,
U_{{\beta}j}^{*}
\right|^2
=
{ 1 \over N }
\sum_{\beta}
\left| A^{e}_{\beta} \right|^2
=
{ 1 \over N }
\;.
\label{15}
\end{equation}
Thus,
in the case of maximal mixing,
if the condition (\ref{101})
is satisfied,
the probability of transitions of
solar $\nu_e$'s into any
state $\nu_\alpha$,
active or sterile,
is equal to $1/N$.
This probability does not depend
on the values
of the amplitudes
$ A^{e}_{\alpha} $,
which means that the matter effect is not observable
in the case of maximal mixing.
In Ref.\cite{HPS96}
this result was obtained for
the cases $N=2$ and $N=3$
with a numerical solution of the evolution
equation of neutrinos in matter.

Let us notice that
the result given in Eq.(\ref{15})
can also be obtained by writing the
neutrino state on the surface of
the sun as a superposition
of mass eigenstates
\begin{equation}
\left| \nu \right\rangle_{\odot}
=
\sum_{j=1}^{N}
A^{e}_{j}
\left| \nu_{j} \right\rangle
\;,
\quad
\mbox{with}
\quad
A^{e}_{j}
=
\sum_{\alpha}
A^{e}_{\alpha}
\,
U_{{\alpha}j}^{*}
\;.
\label{61}
\end{equation}
In general the amplitudes
$A^{e}_{j}$,
which are normalized by
$ \displaystyle
\sum_{j}
\left| A^{e}_{j} \right|^2
=
1
$,
depend on the neutrino energy
and there is more than one
$A^{e}_{j}$ different from zero.
For example,
in the case of three generations of neutrinos
with
$ \displaystyle
( \Delta m^2_{21} / E )
\lesssim
10^{-5} \, \mbox{eV}^2 \, \mbox{MeV}^{-1}
$
an electron neutrino produced
in the core of the sun
is the following superposition\footnote{
This equation
can be obtained from Eqs.(2.66) and (2.69)
of Ref.\cite{KP89}
with
the mixing angles
$ \omega $,
$ \psi$,
$ \varphi $
and the phase $\delta$
such that
$ \sin \omega = \sin \psi = 1 / \sqrt{2} $,
$ \sin \varphi =  1 / \sqrt{3} $
and
$ \delta = \pi/4 $,
which correspond to maximal mixing.
}
of
effective mass eigenstates
$\nu_{2}^{M}$
and
$\nu_{3}^{M}$
\begin{equation}
\left| \nu_{e}^{M} \right\rangle
=
\sqrt{2\over3}
\,
\left| \nu_{2}^{M} \right\rangle
+
\sqrt{1\over3}
\,
\left| \nu_{3}^{M} \right\rangle
\;.
\label{63}
\end{equation}
In this case,
if the propagation of the neutrino
in the interior of the sun is adiabatic,
only
$A^{e}_{2}$ and $A^{e}_{3}$
are different from zero.
If the propagation of the neutrino
in the interior of the sun is non-adiabatic,
transitions from
$\nu_{2}^{M}$
and
$\nu_{3}^{M}$
to
$\nu_{1}^{M}$
are possible
and all three
$A^{e}_{1}$, $A^{e}_{2}$ and $A^{e}_{3}$
are different from zero.
From Eq.(\ref{61}),
for the measurable averaged probability of
$ \nu_e \to \nu_{\alpha} $
transitions,
in which all the interference terms disappear,
we have
\begin{equation}
\left\langle P_{\nu_e\to\nu_\alpha} \right\rangle
=
\sum_{j=1}^{N}
\left| A^{e}_{j} \right|^2
\left| U_{{\alpha}j} \right|^2
=
{1 \over N }
\;.
\label{62}
\end{equation}

It is instructive
to present another derivation of this
result.
If all the oscillating terms in the expression for
the averaged probability of
$ \nu_e \to \nu_{\alpha} $
transitions
disappear,
we have
(see Ref.\cite{KP89,Baldini87})
\begin{equation}
\left\langle P_{\nu_e\to\nu_\alpha} \right\rangle
=
\sum_{j,k}
\left| U_{{\alpha}k} \right|^2
\,
P_{kj}
\,
\left| U_{ej}^{0} \right|^2
\;.
\label{54}
\end{equation}
Here 
$U_{{\alpha}j}^{0}$
are the elements of the mixing matrix 
at the point where
the neutrino is produced and 
$P_{kj}$
is the transition probability
from the state with energy
$E_j$
in the production point to the state
with energy
$E_k$
on the earth.
Using the unitarity relations
\begin{equation}
\sum_{k}
P_{kj}
= 1
\qquad
\mbox{and}
\qquad
\sum_{j}
\left| U_{ej}^{0} \right|^2
=
1
\;,
\label{55}
\end{equation}
in the case of maximal mixing
we obtain Eq.(\ref{53}).

Let us notice that in the case of
maximal mixing
of two massive neutrino fields
there is no MSW
\cite{MSW}
resonance in matter.
In the case of
maximal mixing
of three massive neutrino fields,
if
$ \Delta m^2_{31} $
is large
(say,
$ \Delta m^2_{31} \simeq 10^{-2} \, \mbox{eV}^2 $)
the resonance condition has the form
\begin{equation}
2 \, \sqrt{2} \, G_{F} \, E \, N_{e}
=
\Delta m^2_{21}
\,
{\displaystyle
\cos 2\omega
\over\displaystyle
\cos^2 \varphi
}
\;,
\label{99}
\end{equation}
where $N_{e}$ is the electron density
and the Maiani parameterization of the mixing
matrix has been used
(see Ref.\cite{KP89}).
In the case of maximal mixing
the right-hand side of Eq.(\ref{99})
is equal to zero and there is no MSW resonance in matter.

In Refs.\cite{HPS95,GKK95}
it has been shown that
the atmospheric neutrino data
and
the data of
the Kamiokande, GALLEX and SAGE experiments
\cite{KAMIOKANDE,GALLEX,SAGE}
are well described
in the case of maximal mixing of three neutrinos
with
$ \Delta m^2_{31} \simeq 10^{-2} \, \mbox{eV}^2 $
and
$ \Delta m^2_{21} \lesssim 10^{-12} \, \mbox{eV}^2 $.
In this case the value of
$ P_{\nu_e\to\nu_e} $
does not depend on energy
and is equal to $5/9$.
We will discuss now the possibilities
for the solar neutrino experiments
of the next generation
to check this model.

During 1996 two new solar neutrino
experiments will start,
Super-Kamiokande (S-K) \cite{SK})
and
SNO \cite{SNO}.
In the S-K experiment
solar neutrinos will be detected
through the observation of
elastic scattering (ES)
of neutrinos on electrons,
\begin{equation}
\nu + e^{-} \to \nu + e^{-}
\;.
\label{ES}
\end{equation}

In the SNO experiment
solar neutrinos will be detected
through the observation of the
charged-current (CC)
and
neutral current (NC)
processes
\begin{eqnarray}
&&
\nu_{e} + d \to e^{-} + p + p
\;,
\label{CC}
\\
&&
\nu + d \to \nu + p + n
\;,
\label{NC}
\end{eqnarray}
and also the ES process (\ref{ES}).

In the SNO experiment
the spectrum of electrons in the
CC process (\ref{CC})
will be measured
and the flux
of solar $\nu_e$'s on the earth
as a function of neutrino energy $E$
will be determined
\cite{SNO}.

In both the S-K and SNO experiments,
due to the high energy thresholds
(about 5 MeV for CC and ES and 2.2 MeV for NC)
only neutrinos coming from
$^8\mathrm{B}$ decay
will be detected.
The energy spectrum of the initial $^8\mathrm{B}$ $\nu_e$'s
can be written as
\begin{equation}
\phi_{\mathrm{B}}(E)
=
\Phi_{\mathrm{B}}
\,
X(E)
\;.
\label{500}
\end{equation}
Here $X(E)$ is a known normalized function
determined mainly by the phase space factor
of the decay
$ \mbox{}^8\mathrm{B} \to \mbox{}^8\mathrm{Be} + e^{+} + \nu_{e} $
(see Ref.\cite{Bahcall96}),
and
$\Phi_{\mathrm{B}}$
is the total flux of initial $^8\mathrm{B}$ solar $\nu_{e}$'s.
In the following
we will not make any assumption about
the value of
$\Phi_{\mathrm{B}}$.

In the maximal mixing model under consideration
the survival probability of solar $\nu_e$'s
has the constant value $5/9$
and the shape of the spectrum of $\nu_e$'s
on the earth is given by $X(E)$.
In the high-statistic S-K experiment
the spectrum of recoil electrons
will be measured with high accuracy.
In the model under consideration
the spectrum of the recoil electrons
is given by
\begin{equation}
n^{\mathrm{ES}}(T)
=
{ 5 \over 9 }
\,
\Phi_{\mathrm{B}}
\int_{E_{\mathrm{m}}(T)}
\left( 
{ \mbox{d} \sigma_{\nu_{e}e}
\over
\mbox{d} T }
(E,T)
+
{ 4 \over 5 }
\,
{ \mbox{d} \sigma_{\nu_{\mu}e}
\over
\mbox{d} T }
(E,T)
\right)
X(E)
\mbox{d}E
\;,
\label{505}
\end{equation}
where $T$ is the electron
kinetic energy,
$
\mbox{d} \sigma_{\nu_{\ell}e}
/
\mbox{d} T
$
is the differential cross section
of the ES process
$ \nu_{\ell} e \to \nu_{\ell} e $
(with $\ell=e,\mu$)
and
$
E_{\mathrm{m}}(T)
=
(
1
+
\sqrt{ 1 + 2 \, m_{e} / T }
) \,
T / 2
$.
Let us notice that
the main contribution to
$n^{\mathrm{ES}}(T)$
comes from the first term in the integral.
A comparison of the shape of
the CC and ES spectra
with the SNO and S-K data
will be a test
for the scheme with maximal mixing.
Furthermore,
this model allows
to predict the ratios of different observables
independently from the value of the total
$^{8}\mathrm{B}$ neutrino flux.
In fact,
for the ratio of the total
ES and CC event rates
$
R^{\mathrm{ES}}_{\mathrm{CC}}
=
N^{\mathrm{ES}} / N^{\mathrm{CC}}
$
we have
\begin{equation}
R^{\mathrm{ES}}_{\mathrm{CC}}
=
{\displaystyle
\int_{E_{\mathrm{th}}^{\mathrm{ES}}}
\left( 
\sigma_{\nu_{e}e}(E)
+
{ 4 \over 5 }
\,
\sigma_{\nu_{\mu}e}(E)
\right)
X(E)
\mbox{d}E
\over \displaystyle
\int_{E_{\mathrm{th}}^{\mathrm{CC}}}
\sigma^{\mathrm{CC}}_{\nu_{e}d}(E)
X(E)
\mbox{d}E
}
\;,
\label{506}
\end{equation}
where
$ \sigma_{\nu_{e}d}^{\mathrm{CC}}(E) $
is the cross section for the CC process (\ref{CC})
and
$ E_{\mathrm{th}}^{\mathrm{ES}} $ and
$ E_{\mathrm{th}}^{\mathrm{CC}} $
are the neutrino energy thresholds.
The result of our calculation of
$ R^{\mathrm{ES}}_{\mathrm{CC}} $
is presented in Table \ref{T1},
where it is compared with the corresponding values
calculated in the usual
model with mixing of
two massive neutrino fields
and mixing parameters
$ \Delta m^2 $, $ \sin^2 2\theta $
obtained from the analysis of the
solar neutrino data
(in Ref.\cite{HL}
for MSW transitions
and in Ref.\cite{KP}
for vacuum oscillations).
We used the cross section
$ \sigma_{\nu_{e}d}^{\mathrm{CC}}(E) $
given in
Ref.\cite{KN94}
and an electron kinetic energy threshold
of 4.5 MeV.

Let us consider now the ratio
of the recoil electron spectrum in the ES process
and
the total CC event rate
$
r^{\mathrm{ES}}_{\mathrm{CC}}(T)
=
n^{\mathrm{ES}}(T) / N^{\mathrm{CC}}
$.
We have
\begin{equation}
r^{\mathrm{ES}}_{\mathrm{CC}}(T)
=
{\displaystyle
\int_{E_{\mathrm{m}}(T)}
\left( 
{ \mbox{d} \sigma_{\nu_{e}e}
\over
\mbox{d} T }
(E,T)
+
{ 4 \over 5 }
\,
{ \mbox{d} \sigma_{\nu_{\mu}e}
\over
\mbox{d} T }
(E,T)
\right)
X(E)
\mbox{d}E
\over \displaystyle
\int_{E_{\mathrm{th}}^{\mathrm{CC}}}
\sigma^{\mathrm{CC}}_{\nu_{e}d}(E)
X(E)
\mbox{d}E
}
\;.
\label{507}
\end{equation}
In Fig.\ref{fig1} we have plotted
this ratio as a function of $T$
in the interval
$ 4.5 \, \mbox{MeV} \le T \le 14 \, \mbox{MeV} $.
In Fig.\ref{fig1}
we have also plotted the ratio
$ r^{\mathrm{ES}}_{\mathrm{CC}}(T) $
calculated in the usual model
with mixing of two massive neutrinos
for the cases of MSW transitions
and vacuum oscillations.
We have used the values
of the mixing parameters
$ \Delta m^2 $, $ \sin^2 2\theta $
given in Table \ref{T1}.
It can be seen from Fig.\ref{fig1}
that the investigation of the dependence of
$ r^{\mathrm{ES}}_{\mathrm{CC}}(T) $
on $T$ in the region of small $T$
($ T \lesssim 7 \, \mbox{MeV} $)
will allow to distinguish the case of
maximal mixing of three neutrinos
from the
usual large mixing angle MSW solution
and from vacuum oscillations
of two neutrinos.

For the ratio of the total CC and NC
event rates
$
R^{\mathrm{CC}}_{\mathrm{NC}}
=
N^{\mathrm{CC}} / N^{\mathrm{NC}}
$
we have
\begin{equation}
R^{\mathrm{CC}}_{\mathrm{NC}}
=
{ 5 \over 9 }
\,
{\displaystyle
\int_{E_{\mathrm{th}}^{\mathrm{CC}}}
\sigma^{\mathrm{CC}}_{\nu_{e}d}(E)
X(E)
\mbox{d}E
\over \displaystyle
\int_{E_{\mathrm{th}}^{\mathrm{NC}}}
\sigma^{\mathrm{NC}}_{{\nu}d}(E)
X(E)
\mbox{d}E
}
\;,
\label{503}
\end{equation}
where
$ \sigma^{\mathrm{NC}}_{{\nu}d}(E) $
and
$ E_{\mathrm{th}}^{\mathrm{NC}} = 2.2 \, \mbox{MeV} $ 
are the cross section
and
the energy threshold
of the NC process (\ref{NC}).
The result of our calculation of
$ R^{\mathrm{CC}}_{\mathrm{NC}} $
is presented in Table \ref{T1}.
We used the cross section
$ \sigma^{\mathrm{NC}}_{{\nu}d}(E) $
given in
Ref.\cite{KN94}.

Finally,
for the ratio of
the total
ES and NC event rates
$
R^{\mathrm{ES}}_{\mathrm{NC}}
=
N^{\mathrm{ES}} / N^{\mathrm{NC}}
$
we have
\begin{equation}
R^{\mathrm{ES}}_{\mathrm{NC}}
=
{ 5 \over 9 }
\,
{\displaystyle
\int_{E_{\mathrm{th}}^{\mathrm{ES}}}
\left( 
\sigma_{\nu_{e}e}(E)
+
{ 4 \over 5 }
\,
\sigma_{\nu_{\mu}e}(E)
\right)
X(E)
\mbox{d}E
\over \displaystyle
\int_{E_{\mathrm{th}}^{\mathrm{NC}}}
\sigma^{\mathrm{NC}}_{{\nu}d}(E)
X(E)
\mbox{d}E
}
\;.
\label{508}
\end{equation}
The result of our calculation
of
$ R^{\mathrm{ES}}_{\mathrm{NC}} $
is presented in Table \ref{T1}.
It can be seen from Table \ref{T1}
that the measurement of the ratios
$ R^{\mathrm{ES}}_{\mathrm{CC}} $,
$ R^{\mathrm{CC}}_{\mathrm{NC}} $
and
$ R^{\mathrm{ES}}_{\mathrm{NC}} $
will allow to distinguish the case
of maximal mixing of three neutrinos
from all the usual two-neutrinos
solutions of the solar neutrino problem.

The comparison of the relations
(\ref{505})--(\ref{508})
with the data of the S-K and SNO
experiments
will be a crucial test of the
maximal mixing model
of Refs.\cite{HPS95,GKK95}.

In conclusion,
we have presented simple and general arguments
based on the unitarity
of the mixing matrix
which show that
matter effects do not contribute
to the values of the averaged transition
probabilities of solar $\nu_e$'s
into other states
in the case of maximal mixing.
We have also shown
that the future
solar neutrino experiments
(Super-Kamiokande and SNO),
in which solar $^{8}\mbox{B}$ neutrinos
will be detected through
CC and NC reactions,
will allow to check
the model with maximal mixing
of three massive neutrino fields
that describes the
atmospheric neutrino anomaly
and the data of the Kamiokande,
GALLEX and SAGE experiments. 

\acknowledgments

We would like to thank Serguey Petcov
for useful discussions.
S.B. would like to acknowledge
the kind hospitality of the
Department of Theoretical Physics
of the University of Torino
and
the Elementary Particle Sector of SISSA,
where this work has been done.

\widetext

\begin{table}
\protect\caption{
Results of the calculation of the ratios
$ R^{\mathrm{ES}}_{\mathrm{CC}} $,
$ R^{\mathrm{CC}}_{\mathrm{NC}} $
and
$ R^{\mathrm{ES}}_{\mathrm{NC}} $
in the model with
maximal mixing of three massive neutrino fields
(see Eqs.(\protect\ref{506}),
(\protect\ref{503}) and (\protect\ref{508})).
The values of these ratio
calculated in the usual model with mixing of
two massive neutrinos
are also presented.
The values of the
mixing parameters
$ \Delta m^2 $, $ \sin^2 2\theta $
have been obtained from the analysis of the
solar neutrino data
in Ref.\protect\cite{HL}
for the case of MSW transitions
and in Ref.\protect\cite{KP}
for the case of vacuum oscillations.
}
\begin{tabular}{cccccc}
\\
$ \nu_{e} \to \nu_{\mu(\tau)} $
&
$ \Delta m^2 \, (\mathrm{eV}^2) $
&
$ \sin^2 2\theta $
&
$ R^{\mathrm{ES}}_{\mathrm{CC}} $
&
$ R^{\mathrm{CC}}_{\mathrm{NC}} $
&
$ R^{\mathrm{ES}}_{\mathrm{NC}} $
\\
\\
\hline
\\
MAXIMAL MIXING
&
&
&
$ 2.1 \times 10^{-2} $
&
1.33
&
$ 2.8 \times 10^{2} $
\\
\\
SMALL MIX. MSW
&
$ 6.1 \times 10^{-6} $
&
$ 6.5 \times 10^{-3} $
&
$ 2.4 \times 10^{-2} $
&
0.79
&
$ 1.9 \times 10^{2} $
\\
\\
LARGE MIX. MSW
&
$ 9.4 \times 10^{-6} $
&
$ 0.62 $
&
$ 3.1 \times 10^{-2} $
&
0.46
&
$ 1.4 \times 10^{2} $
\\
\\
VACUUM OSC.
&
$ 8.0 \times 10^{-11} $
&
$ 0.80 $
&
$ 2.7 \times 10^{-2} $
&
0.68
&
$ 1.9 \times 10^{2} $
\\
\\
\end{tabular}
\label{T1}
\end{table}

\narrowtext

\begin{figure}[h]
\protect\caption{
Results of the calculation of the ratio
$ r^{\mathrm{ES}}_{\mathrm{CC}}(T) $
as a function of
the electron recoil kinetic energy $T$
(see Eq.(\protect\ref{507})).
The curves of
$ r^{\mathrm{ES}}_{\mathrm{CC}}(T) $
calculated in the usual model with mixing of
two massive neutrinos
are also presented.
The values of the mixing parameters
$ \Delta m^2 $, $ \sin^2 2\theta $
are given in Table \protect\ref{T1}.
}
\label{fig1}
\end{figure}

%\end{document}

\newpage

\begin{minipage}[h]{\textwidth}
\null\vskip-5cm
\begin{center}
\mbox{\epsfig{file=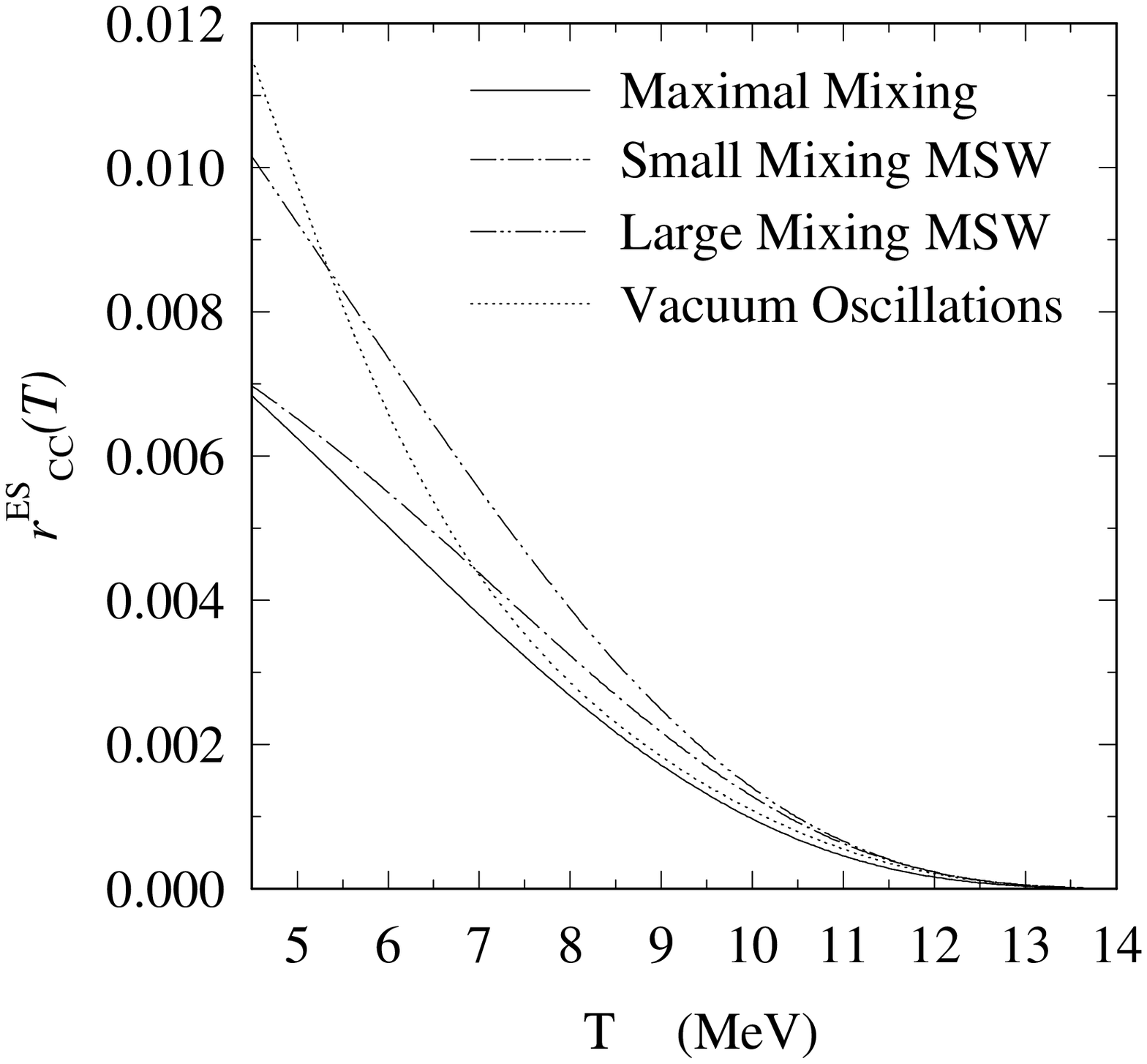,width=\textwidth}}
\end{center}
\end{minipage}
\vspace{1cm}
\begin{center}
{\Large Figure \ref{fig1}}
\end{center}

\end{document}